\documentclass[aps,twocolumn,showpacs,preprintnumbers,floats]{revtex4}\def\@cite#1#2{\textsuperscript{[{#1\if@tempswa , #2\fi}]}}

\usepackage{graphicx}
\usepackage{amsmath}
\usepackage{amsfonts}
\usepackage{amssymb}
\usepackage{color}
\usepackage{subfigure}
\usepackage{epsfig}
\usepackage{morefloats}
\usepackage{multirow}
\usepackage{graphicx,booktabs}
\usepackage{mathrsfs}
\usepackage{txfonts}
\usepackage{indentfirst}
\usepackage{graphicx,booktabs}

\usepackage{longtable,lscape}
\usepackage[colorlinks, citecolor=blue,anchorcolor=red,menucolor=red, linkcolor=red,filecolor=red,urlcolor=blue,frenchlinks=red]{hyperref}

\usepackage{comment}

\begin{document}


\title{Toward establishing the low-lying $P$-wave $\Sigma_b$ states}

\author{Li-Ye Xiao$^{1,2}$~\footnote {E-mail: lyxiao@ustb.edu.cn} and Xian-Hui Zhong$^{2,3}$~\footnote {E-mail: zhongxh@hunnu.edu.cn}}

\affiliation{ 1)School of Mathematics and Physics, University of Science and Technology Beijing,
Beijing 100083, China}
\affiliation{ 2) Department of
Physics, Hunan Normal University, and Key Laboratory of
Low-Dimensional Quantum Structures and Quantum Control of Ministry
of Education, Changsha 410081, China }
\affiliation{ 3) Synergetic
Innovation Center for Quantum Effects and Applications (SICQEA),
Hunan Normal University, Changsha 410081, China}


\begin{abstract}

In the present work, we analyze the $P$-wave singly-heavy baryon
spectrum belonging to $\mathbf{6}_F$ by combining the observations of
the heavy baryon states, and restudy the strong decays of the $1P$ wave $\Sigma_b$
states within the $j$-$j$ coupling scheme using the chiral quark model.
We obtain that: (i) the structure $\Sigma_b(6097)$ observed in the  $\Lambda_b\pi$ final state may arise from the overlapping of $\Sigma_b|J^P=\frac{3}{2}^-,2\rangle$ and $\Sigma_b|J^P=\frac{5}{2}^-,2\rangle$. (ii) The broad structure $\Sigma_b(6072)$ observed in the  $\Lambda_b\pi\pi$ final state may arise from the overlapping of $\Sigma_b|J^P=\frac{1}{2}^-,1\rangle$ and $\Sigma_b|J^P=\frac{3}{2}^-,1\rangle$. (iii) The missing state $\Sigma_b|J^P=\frac{1}{2}^-,0\rangle$ is most likely to be a narrow state with a
width of $\Gamma\sim10$ MeV, and mainly decays into $\Lambda_b\pi$ channel.

\end{abstract}


\maketitle

\section{Introduction}

The LHC experiments have demonstrated their discovery capability of heavy flavored baryons.
During the past several years, many missing heavy baryon states have been
discovered by LHC experiments. For the charmed baryon sector,
in 2017 the LHCb Collaboration observed five extremely narrow $\Omega_c(X)$
states, $\Omega_c(3000)$, $\Omega_c(3050)$, $\Omega_c(3066)$, $\Omega_c(3090)$ and $\Omega_c(3119)$,
in the $\Xi_c^{+}K^-$ channel~\cite{Aaij:2017nav},
and the first four of them were confirmed by the subsequent Belle experiments~\cite{Yelton:2017qxg}.
Very recently, the LHCb Collaboration observed three new states $\Xi_c(2923)^0$, $\Xi_c(2939)^0$, and $\Xi_c(2965)^0$ in
the $\Lambda_c^+K^-$ mass spectrum with a large significance~\cite{Aaij:2020yyt}.
For the bottom baryon sector, a great progress has been achieved as well.
In 2018, the LHCb collaboration reported two new bottom baryon states $\Sigma_b(6097)^{\pm}$
in the $\Lambda_b^0 \pi^{\pm}$ channels~\cite{Aaij:2018tnn}, and another new state $\Xi_b(6227)^-$ in both
$\Lambda_b^0K^-$ and $\Xi_b^0\pi$ decay modes~\cite{Aaij:2018yqz}.
Very recently, the LHCb Collaboration observed four narrow $\Omega_b(X)$ states
$\Omega_b(6316)$, $\Omega_b(6330)$, $\Omega_b(6340)$, and $\Omega_b(6350)$~\cite{Aaij:2020cex},
in the $\Xi^0_bK^-$ final state.

\begin{table*}[ht]
\caption{The quark model classifications of these newly observed resonances based on our previous works\cite{Wang:2020gkn,Xiao:2020oif,Wang:2018fjm,Wang:2017kfr,Wang:2017hej}.} \label{classfy}
\begin{tabular}{cccccccccccccccc }\hline\hline
$L$-$S$ scheme& $j$-$j$ scheme & \multicolumn{6}{c}{Observed states/structures belonging to $\mathbf{6}_F$ multiplet.} \\ \hline
~~~~$|n^{2S+1}L_{\lambda}J^P\rangle$~~~~~~& ~~~~$|J^P,j\rangle~(nl)$  ~~~~~~~~~~~~&$\Omega_c$ states~~~~~~&$\Xi_c'$ states~~~~~~&$\Sigma_c$ states
~~~~~~&$\Omega_b$ states~~~~~~&$\Xi_b'$ states~~~~~~&$\Sigma_b$ states \\
\hline
$|1^{2}S \frac{1}{2}^+\rangle$ &$|J^P=\frac{1}{2}^+,1\rangle(1S)$~~&$\Omega_c(2695)$ ~~&$\Xi_c'(2578)$~~&$\Sigma_c(2455)$~~&$\Omega_b(6046)$~~&$\Xi_b'(5935)$~~&$\Sigma_b(5810)$\\
$|1^{4}S \frac{3}{2}^+\rangle$ &$|J^P=\frac{3}{2}^+,1\rangle(1S)$~~&$\Omega_c^*(2770)$ ~~&$\Xi_c^*(2645)$~~&$\Sigma_c^*(2520)$~~&$\cdots$~~&$\Xi_b^*(5955)$~~&$\Sigma_b^*(5830)$\\
$|1P_\lambda \frac{1}{2}^-\rangle_1$ &$|J^P=\frac{1}{2}^-,1\rangle(1P)$~~&$\Omega_c(3000)$ ~~&$\cdots$~~&$\cdots$~~&$\Omega_b(6316)$~~&$\cdots$~~&$\Sigma_b(6072)$  \\
$|1P_\lambda \frac{1}{2}^-\rangle_2$ &$|J^P=\frac{1}{2}^-,0\rangle(1P)$~~&$\cdots$ ~~&$\Xi_c(2880)$~~&$\cdots$~~&$\cdots$~~&$\cdots$~~&$\cdots$~~\\
$|1^{4}P_\lambda \frac{3}{2}^-\rangle$ &$|J^P=\frac{3}{2}^-,1\rangle(1P)$~~&$\Omega_c(3050)$ ~~&$\Xi_c(2923)$~~&$\cdots$~~&$\Omega_b(6330)$~~&$\cdots$~~&$\Sigma_b(6072)$ \\
$|1^{2}P_\lambda \frac{3}{2}^-\rangle$ &$|J^P=\frac{3}{2}^-,2\rangle(1P)$~~&$\Omega_c(3065)$ ~~&$\Xi_c(2939)$~~&$\Sigma_c(2800)$~~&$\Omega_b(6340)$~~&$\Xi_b'(6227)$~~&$\Sigma_b(6097)$ \\
$|1^{4}P_\lambda \frac{5}{2}^-\rangle$ &$|J^P=\frac{5}{2}^-,2\rangle(1P)$~~&$\Omega_c(3090)$ ~~&$\Xi_c(2965)$~~&$\Sigma_c(2800)$~~&$\Omega_b(6350)$~~&$\Xi_b'(6227)$~~&$\Sigma_b(6097)$ \\
\hline\hline
\end{tabular}
\end{table*}

These newly observed heavy baryon states may be good candidates of the $P$-wave heavy baryons belonging to
the flavor sextet $\mathbf{6}_F$, when compared with the predicted masses from virous phenomenological models and methods~\cite{Capstick:1986bm,Capstick:2000qj,Ebert:2011kk,Ebert:2007nw,Ebert:2005xj,Yoshida:2015tia,Roberts:2007ni,Valcarce:2008dr,Karliner:2008sv,Chen:2014nyo,Mao:2015gya,
Wang:2010it,Thakkar:2016dna,Yang:2017qan,Garcilazo:2007eh,Wei:2016jyk,Jia:2019bkr,Yamaguchi:2014era,Chen:2018vuc}.
In our previous works~\cite{Wang:2020gkn,Xiao:2020oif,Wang:2018fjm,Wang:2017kfr,Wang:2017hej},
we discussed the quark model classifications of these
newly observed resonances. Our results are summarized in Table~\ref{classfy}.
It is found that a relatively complete $\lambda$-mode $P$-wave heavy baryon spectrum may have been established
with discovering the series of heavy baryons.

Recently, the CMS collaboration observed a broad enhancement around 6070 MeV in the
$\Lambda^0_b\pi^+\pi^-$ invariant mass spectrum~\cite{Sirunyan:2020gtz}, which was confirmed
by the subsequent LHCb experiments with high significance~\cite{Aaij:2020rkw}.
The mass and width for the structure are determined to be
\begin{eqnarray}
M=6072.3\pm2.9\pm0.6\pm0.2~\text{MeV},\\
\Gamma=72\pm11\pm2~\text{MeV}.
\end{eqnarray}
From the point of view of mass, this structure around 6070 MeV [denoted with $\Sigma_b(6072)$]
may be some signals of the $P$-wave states in the $\Sigma_b$ family, although
it was interpreted as the first radially excited $\Lambda_b$ state, $\Lambda_b(2S)$,
in the literature~\cite{Aaij:2020rkw,Arifi:2020yfp}. Together with the other
$P$-wave candidate $\Sigma_b(6097)$ observed in the $\Lambda_b^0 \pi^{\pm}$
channels about two years ago at LHCb~\cite{Aaij:2018tnn}, the $\Sigma_b(6072)$ provide us
a good chance to establish a fairly complete $P$-wave spectrum in
the $\Sigma_b$ family.

In the present work we re-study the $1P$ wave $\Sigma_b$ states. We hope to understand the nature of
the broad structure $\Sigma_b(6072)$ observed by the CMS collaboration~\cite{Sirunyan:2020gtz} and
LHCb collaboration~\cite{Aaij:2020rkw}. Furthermore, by combining the newest observations
of the heavy baryon states, we give our opinion for establishing a complete $P$-wave
$\Sigma_b$ spectrum.

This paper is organized as follows. In Sec. II we give an analysis of the $1P$ wave singly-heavy baryons' spectra. We discuss the strong decay properites of the $1P$ wave $\Sigma_b$ states within the $j$-$j$ coupling scheme in Sec. III and summarize our results in Sec. IV.

\begin{table*}[ht]
\caption{The predicted masses (MeV) of the $\Sigma_b^+$ states within various phenomenal models and methods. $|J^P,j\rangle~(nl)$ denotes the state within the $j$-$j$ coupling scheme.} \label{mass-spectrum}
\begin{tabular}{cccccccccccccccc }\hline\hline
$|J^P,j\rangle~(nl)$ ~~~&RQM~~\cite{Capstick:1986bm} ~~&hCQM~\cite{Thakkar:2016dna}~~&RQM~\cite{Ebert:2011kk}
~~&ChQM~\cite{Yang:2017qan}~~&QM~\cite{Roberts:2007ni}~~&CQM~\cite{Garcilazo:2007eh}~~&CQM~\cite{Ebert:2007nw} ~~&Observed state \\
\hline
$|J^P=\frac{1}{2}^+,1\rangle(1S)$~~&5795 ~~&5811~~&5808~~&5812~~&5833~~&5789~~&5805 ~~&$\Sigma_b^+$ \\
$|J^P=\frac{3}{2}^+,1\rangle(1S)$~~&5805 ~~&5832~~&5834~~&5824~~&5858~~&5844~~&5834 ~~&$\Sigma_b^{*+}$\\
$|J^P=\frac{1}{2}^-,1\rangle(1P)$~~&6070 ~~&6139~~&6101~~&6039~~&6099~~&6039~~&6122 ~~& $\Sigma_b(6072)$ \\
$|J^P=\frac{1}{2}^-,0\rangle(1P)$~~&6070 ~~&6148~~&6095~~&6039~~&6106~~&6142~~&6108 ~~& $\cdots$\\
$|J^P=\frac{3}{2}^-,2\rangle(1P)$~~&6070 ~~&6120~~&6096~~&6039~~&6101~~&6039~~&6106 ~~&$\Sigma_b(6097)$ \\
$|J^P=\frac{3}{2}^-,1\rangle(1P)$~~&6085 ~~&6129~~&6087~~&6039~~&6105~~&6142~~&6076 ~~& $\Sigma_b(6072)$\\
$|J^P=\frac{5}{2}^-,2\rangle(1P)$~~&6090 ~~&6104~~&6084~~&6039~~&6172~~&$\cdot\cdot\cdot$ ~~&6083 ~~&$\Sigma_b(6097)$ \\
\hline\hline
\end{tabular}
\end{table*}

\begin{figure*}[]
\centering \epsfxsize=17 cm \epsfbox{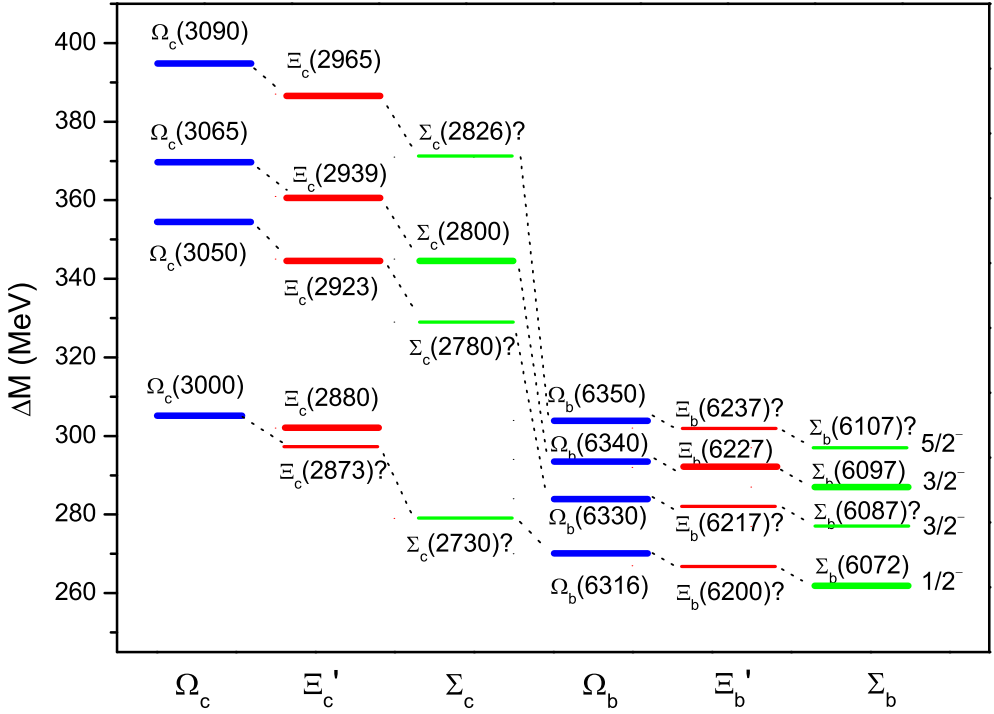} \vspace{-1.2cm} \caption{The
mass splitting ($\Delta M$) between the newly observed heavy baryons and their ground states.
The states labeled with $``?"$ are predictions in this work.}\label{figmasssplit}
\end{figure*}


\section{mass spectrum analysis}\label{massspecrum}

In the $L$-$S$ coupling scheme, there are five $P$-wave states,
$|1^{2}P_\lambda \frac{1}{2}^-\rangle$, $|1^{2}P_\lambda \frac{3}{2}^-\rangle$,
$|1^{4}P_\lambda \frac{1}{2}^-\rangle$, $|1^{4}P_\lambda \frac{3}{2}^-\rangle$ and $|1^{4}P_\lambda \frac{5}{2}^-\rangle$, for a flavor  multiplet $\mathbf{6}_F$. The physical states may be closer to the $j$-$j$
coupling scheme due to the heavy quark symmetry. In this coupling scheme,
there five $P$-wave states are denoted as $|J^P=\frac{1}{2}^-,0\rangle $, $|J^P=\frac{1}{2}^-,1\rangle $,
$|J^P=\frac{3}{2}^-,1\rangle $, $|J^P=\frac{3}{2}^-,2\rangle $, and $|J^P=\frac{5}{2}^-,2\rangle $.
The two $J^P=1/2^-$ states $|J^P=\frac{1}{2}^-,0\rangle$ and $|J^P=\frac{1}{2}^-,1\rangle$ in
the $j$-$j$ scheme are mixed states between $|^2P_{1/2}\rangle$ and $|^4P_{1/2}\rangle$ of the $L$-$S$ coupling scheme with a mixing angle
$\phi\simeq35^\circ$. While the two $J^P=3/2^-$ states $|J^P=\frac{3}{2}^-,1\rangle $ and $|J^P=\frac{3}{2}^-,2\rangle $ are mixed states via a $|^2P_{3/2}\rangle$-$|^4P_{3/2}\rangle$ mixing with a relatively small mixing angle $\phi\simeq24^\circ$.

It is interesting to find that the newly observed state $\Omega_c(3000)$ at LHCb~\cite{Aaij:2017nav} may be explained as a $J^P=1/2^-$ resonance via a $|^2P_{1/2}\rangle$-$|^4P_{1/2}\rangle$ mixing with a mixing angle $\phi\simeq24^\circ$~\cite{Wang:2017hej},
which is smaller than $35^\circ$ from the $j$-$j$ coupling scheme. This may indicate that the physical states of the
singly-heavy baryon states should lie between $L$-$S$ and $j$-$j$ coupling schemes.
The physical state for the bottom sector may be closer to the $j$-$j$ coupling scheme than that for the charmed sector
because the bottom quark is much heavy than the charm quark.

It is difficult to predict the mass order and mass splittings in theory due to the
unclear spin-dependent interactions for quarks.
In fact, the mass order and mass splittings between the $P$-wave spin multiplets
are crucial to determine their quantum numbers. Fortunately, the recent observations of the heavy baryons in
the $\Omega_c$, $\Xi_c'$, and $\Omega_b$ families provide us important information about the mass order and mass splitting.
For clarity, we further plot the mass splittings between the
newly observed heavy baryons and their ground states in Figure~\ref{figmasssplit}.
Combining the equal spacing rule predicted in~\cite{GellMann:1962xb,Okubo:1961jc},
we predict that for the charmed sector the flavor partners of the four $\Omega_c(X)$ states
$\Omega_c(3000)$, $\Omega_c(3050)$, $\Omega_c(3065)$ and $\Omega_c(3090)$ are most likely to be
\begin{eqnarray}
\Omega_c(3000)&:& \Xi_{c}(2873)?, \Sigma_{c}(2730)?, \nonumber \\
\Omega_c(3050)&:& \Xi_{c}(2923),\  \Sigma_{c}(2780)?,\nonumber\\
\Omega_c(3065)&:& \Xi_{c}(2939),\  \Sigma_{c}(2800),\nonumber\\
\Omega_c(3090)&:& \Xi_{c}(2965),\  \Sigma_{c}(2826)?,\nonumber
\end{eqnarray}
respectively.
While for the for the bottom sector, the flavor partners of the four $\Omega_b(X)$ states
$\Omega_b(6316)$, $\Omega_b(3330)$, $\Omega_b(6340)$ and $\Omega_b(6350)$ are predicted to be
\begin{eqnarray}
\Omega_b(6316)&:& \Xi_{b}(6203)?, \  \Sigma_{b}(6072), \nonumber \\
\Omega_b(6330)&:& \Xi_{b}(6217)?,\  \Sigma_{b}(6087)?,\nonumber\\
\Omega_b(6340)&:& \Xi_{b}(6227),\  \ \ \Sigma_{b}(6097),\nonumber\\
\Omega_b(6350)&:& \Xi_{b}(6237)?,\  \Sigma_{b}(6107)?,\nonumber
\end{eqnarray}
respectively. The states labeled with $``?"$ are waiting to be discovered
in future experiments.

According to our previous studies in Refs.~\cite{Wang:2017hej,Wang:2020gkn} we further
find that the four $\Omega_b(6316)$, $\Omega_b(3330)$, $\Omega_b(6340)$
and $\Omega_b(6350)$ are flavor partners of the four $\Omega_c(X)$ states
$\Omega_c(3000)$, $\Omega_c(3050)$, $\Omega_c(3065)$ and $\Omega_c(3090)$, respectively.
They are good candidates of the $P$-wave singly-heavy baryon states with
spin-parity numbers $J^P=1/2^-$, $J^P=3/2^-$, $J^P=3/2^-$, $J^P=5/2^-$,
respectively. The possible quark model classifications of
these possible $P$-wave singly-heavy baryons observed in experiments are summarized in Table~\ref{classfy}.
If our assignments are correct, in the heavy quark symmetry limit,
the mass order for the $P$-wave singly-heavy baryon states might be
\begin{eqnarray}
M(|J^P=5/2^-,2\rangle) > M(|J^P=3/2^-,2\rangle)\ \ \ \ \ \ \ \ \ \ \ \ \ \ \nonumber \\
 \ \ \ \ \ \ \ \ \ \ \ \ \ \ > M(|J^P=3/2^-,1\rangle)> M(|J^P=1/2^-,1\rangle).
\end{eqnarray}

Until now, we may conclude that the $\Sigma_b(6097)$ and $\Sigma_b(6072)$, as the flavor partners of
$\Omega_b(6340)$ and $\Omega_b(6316)$, may favor the assignments $|J^P=\frac{3}{2}^-,2\rangle$ and $|J^P=\frac{1}{2}^-,1\rangle$, respectively, according to our classifications of
the newly observed $\Sigma_b(X)$ states. The other two missing $P$-wave $\Sigma_b$
states $\Sigma_b(6087)$ and $\Sigma_b(6107)$, as flavor partners of $\Omega_b(6330)$ and $\Omega_b(6350)$, may correspond to the states $|J^P=\frac{3}{2}^-,1\rangle$ and $|J^P=\frac{5}{2}^-,2\rangle$.
Furthermore, we also should mention that the $\Sigma_b(6097)$ and $\Sigma_b(6072)$
observed in experiments may be caused by several nearby states due to a rather
small mass splitting, $\sim10$ MeV, between two nearby states.
Finally, it should be pointed out that there is less knowledge about the $P$-wave state
$|J^P=\frac{1}{2}^-,0\rangle$. Only a hint comes from the recent LHCb experiments~\cite{Aaij:2020yyt}.
It is found a broad structure $\Xi_c(2880)$ in the $\Lambda_c^+K^-$ mass spectrum with
a small significance~\cite{Aaij:2020yyt}. The $\Xi_c(2880)$ may be a candidate of the
$|J^P=\frac{1}{2}^-,0\rangle$ in the $\Xi_c'$ family. If this is confirmed by future experiments,
the two $P$-wave $J^P=1/2^-$ states $|J^P=\frac{1}{2}^-,1\rangle$ and
$|J^P=\frac{1}{2}^-,0\rangle$ may be largely overlapping states according to
the equal spacing rule~\cite{GellMann:1962xb,Okubo:1961jc}.

As a whole, combining the equal spacing rule with the observations of
the heavy baryon states, we can predict a fairly complete $P$-wave singly-heavy baryon
spectrum belonging to $\mathbf{6}_F$. These predictions may be helpful to looking for the missing
state in forthcoming experiments at LHC and/or Belle II.

\section{strong decay analysis}

As a possible explanation of the broad structure below 6100 MeV in the
$\Lambda_b\pi\pi$ invariant mass spectrum observed by the CMS collaboration~\cite{Sirunyan:2020gtz}
and LHCb collaboration~\cite{Aaij:2020rkw}, it is crucial to re-study the strong decay properties of the
low-lying $\lambda$-mode $1P$ wave $\Sigma_b$ states.
In the following we will analyze the strong decays of the $P$-wave $\Sigma_b$ states with the chiral quark model.
This model has been successfully applied to the strong decays of heavy-light mesons, charmed
and strange baryons~\cite{Zhong:2008kd,Zhong:2010vq,Zhong:2009sk,
Zhong:2007gp,Liu:2012sj,Xiao:2013xi,Wang:2017hej,Xiao:2014ura,Xiao:2017udy,Yao:2018jmc}.
The model parameters have been well determined in theses works. In present work,
the parameters are adopted the same as those of our previous studies of the singly-heavy baryons~\cite{Zhong:2007gp,Wang:2017hej,Xiao:2014ura,Xiao:2017udy,Yao:2018jmc}, where
the details for the framework can be found as well.



\subsection{$J^P=1/2^-$ states}

In previous mass spectrum analysis, it is found that the $\Sigma_b(6072)$ may be assigned as the $J^P=1/2^-$ states $\Sigma_b|J^P=\frac{1}{2}^-,1\rangle $. We further test this assignment from the sector of strong decay properties.

\begin{figure}[]
\centering \epsfxsize=7.5cm \epsfbox{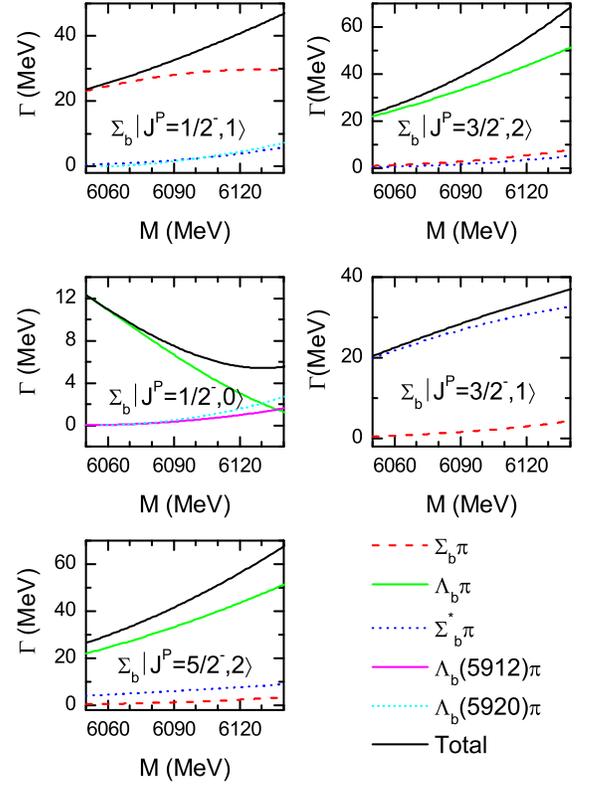} \caption{The
strong decay properties of the $\lambda$-mode $1P$-wave $\Sigma_b$ states as functions of the masses. }\label{fig-mass}
\end{figure}

There are two $J^P=1/2^-$ states $\Sigma_b|J^P=\frac{1}{2}^-,1\rangle $ and
$\Sigma_b|J^P=\frac{1}{2}^-,0\rangle $ within the $j$-$j$ coupling scheme (see Table~\ref{mass-spectrum}).
They can be expressed as mixed states between $|^2P_{\lambda}\frac{1}{2}^-\rangle$ and
$|^4P_{\lambda}\frac{1}{2}^-\rangle$ with the $L$-$S$ coupling scheme:
\begin{eqnarray}
\left|J^P=\frac{1}{2}^-,1\right\rangle=+\sqrt{\frac{2}{3}}\left|^2P_{\lambda}\frac{1}{2}^-\right\rangle+\sqrt{\frac{1}{3}}\left|^4P_{\lambda}\frac{1}{2}^-\right\rangle,\\
\left|J^P=\frac{1}{2}^-,0\right\rangle=-\sqrt{\frac{1}{3}}\left|^2P_{\lambda}\frac{1}{2}^-\right\rangle+\sqrt{\frac{2}{3}}\left|^4P_{\lambda}\frac{1}{2}^-\right\rangle.
\end{eqnarray}
Their masses are predicted to about $M\simeq(6050-6140)$ MeV (see Table~\ref{mass-spectrum}), and their OZI-allowed two body strong decay channels are $\Sigma^{(*)}_b\pi$, $\Lambda_b\pi$, $\Lambda_b(5912)\pi$ and $\Lambda_b(5920)\pi$. To know about the effects of the uncertainties of the masses on the decay properties of the $\Sigma_b|J^P=\frac{1}{2}^-,1\rangle $ and $\Sigma_b|J^P=\frac{1}{2}^-,0\rangle$ states, we plot the partial decay widths as functions of the masses in the range of $M=(6050-6140)$ MeV in Fig.~\ref{fig-mass}.
From the figure, it is found that $\Sigma_b|J^P=\frac{1}{2}^-,1\rangle$ has a total decay width of $\Gamma\sim(24-47)$ MeV with the mass varied in the range we considered, and dominantly decays into the $\Sigma_b\pi$ channel. Meanwhile, the $\Lambda_b\pi$ decay mode is forbidden for $\Sigma_b|J^P=\frac{1}{2}^-,1\rangle$.

Considering the $\Sigma_b(6072)$ observed in the $\Lambda_b\pi^+ \pi^-$ final state at CMS~\cite{Sirunyan:2020gtz} and LHCb~\cite{Aaij:2020rkw}
as the $\Sigma_b|J^P=\frac{1}{2}^-,1\rangle$ state,
its width is predicted to be
\begin{eqnarray}
\Gamma \simeq28~\text{MeV}.
\end{eqnarray}
The predicted branching fraction of the dominant decay mode $\Sigma_b\pi$ is
\begin{eqnarray}
\mathcal{B}[\Sigma_b(6072)\to \Sigma_b\pi]\simeq 93\%.
\end{eqnarray}
The observed $\Lambda_b\pi \pi$ decay mode of $\Sigma_b(6072)$ can
be naturally explain with the cascade decay process $\Sigma_b(6072)\to \Sigma_b\pi\to \Lambda_b\pi\pi$.
The decay rates into the $\Sigma_b^*\pi$, $\Lambda_b(5912)\pi$ and $\Lambda_b(5920)\pi$ final states are small (see Table~\ref{decaywidth}).
The decay predicted width of $\Sigma_b(6072)$ is about a factor 2 smaller than the observations from the CMS~\cite{Sirunyan:2020gtz} and LHCb~\cite{Aaij:2020rkw}, which indicates that the structure $\Sigma_b(6072)^0$ observed at
CMS~\cite{Sirunyan:2020gtz} and LHCb~\cite{Aaij:2020rkw} may not be well understood with the
$\Sigma_b|J^P=\frac{1}{2}^-,1\rangle$ state only.

\begin{table*}[htp]
\begin{center}
\caption{\label{decaywidth}  The decay properties of the $\lambda$-mode $1P$ wave $\Sigma_b$ states. The units of mass and width are MeV in the table.}
\begin{tabular}{cccccccccccccccccccccccccccccccccccccccccccccc}\hline\hline
State~~~~~~ &Mass  ~~~~~~&$\Gamma[\Sigma_b\pi]$  ~~~~~~&$\Gamma[\Lambda_b\pi]$   ~~~~~~&$\Gamma[\Sigma^*_b\pi]$  ~~~~~~&$\Gamma[\Lambda_b(5912)\pi]$  ~~~~~~&$\Gamma[\Lambda_b(5920)\pi]$ ~~~~~~&$\Gamma[\text{total}]$       \\
 \hline
$|J^P=\frac{1}{2}^-, 1\rangle$~~~~~~&6072~~~~~~& 26.2~~~~~~&$\cdot\cdot\cdot$~~~~~~&1.11~~~~~~&0.42~~~~~~&0.44~~~~~~&28.2\\
$|J^P=\frac{1}{2}^-, 0\rangle$~~~~~~&6076~~~~~~&$\cdot\cdot\cdot$~~~~~~&8.75~~~~~~&$\cdot\cdot\cdot$~~~~~~&0.15~~~~~~&0.17~~~~~~&9.08 \\
$|J^P=\frac{3}{2}^-, 2\rangle$~~~~~~&6097~~~~~~& 3.26~~~~~~&35.2~~~~~~&1.97~~~~~~&0.50~~~~~~&0.75~~~~~~&41.7\\
$|J^P=\frac{3}{2}^-, 1\rangle$~~~~~~&6087~~~~~~& 1.46~~~~~~&$\cdot\cdot\cdot$~~~~~~&26.4~~~~~~&$\cdot\cdot\cdot$~~~~~~&$\cdot\cdot\cdot$~~~~~~&27.8 \\
$|J^P=\frac{5}{2}^-, 2\rangle$~~~~~~&6107~~~~~~& 1.84~~~~~~&38.9~~~~~~&7.00 ~~~~~~&0.71~~~~~~&1.12~~~~~~&49.5 \\
\hline\hline
\end{tabular}
\end{center}
\end{table*}

The other $J^P=1/2^-$ state $\Sigma_b|J^P=\frac{1}{2}^-,0\rangle$ should be a very narrow state
with a width of $\Gamma\sim(12-6)$ MeV with the mass varied in the range of $M=(6040-6140)$ MeV.
According to our previous analysis in Sec.~\ref{massspecrum}, the mass of $\Sigma_b|J^P=\frac{1}{2}^-,0\rangle$
may be very close to that of $\Sigma_b(6072)$.
Fixing the mass of $\Sigma_b|J^P=\frac{1}{2}^-,0\rangle$ at $M=6072$ MeV, we obtain
\begin{eqnarray}
\Gamma\simeq9~\text{MeV},
\end{eqnarray}
and the $\Lambda_b\pi$ decay channel almost saturates its width (see Table~\ref{decaywidth}).
The decay properties of this state is inconsistent with the nature of the broad structure
around $M\sim6072$ MeV in the $\Lambda_b\pi\pi$ invariant mass spectrum.
The $\Sigma_b|J^P=\frac{1}{2}^-,0\rangle$ may have large potentials to be
observed in the $\Lambda_b\pi$ final state in forthcoming experiments.
It may be a pity that the narrow state $\Sigma_b|J^P=\frac{1}{2}^-,0\rangle$
was not observed in the previous LHCb experiments~\cite{Aaij:2018tnn}, which may be due to a too large
energy scanned step, $\sim 6$ MeV, adopted in measurements.

\subsection{$J^P=3/2^-$ states}

In previous mass spectrum analysis, it is found that the $\Sigma_b(6097)$ may be assigned as the $J^P=3/2^-$ states $\Sigma_b|J^P=\frac{3}{2}^-,2\rangle $. We further test this assignment from the sector of strong decay properties.

In the $\Sigma_b$ family, there are two $\lambda$-mode $1P$ wave $J^P=3/2^-$ states $\Sigma_b|J^P=\frac{3}{2}^-,2\rangle$ and $\Sigma_b|J^P=\frac{3}{2}^-,1\rangle$. They can relate to the states in the $L$-$S$ coupling scheme:
\begin{eqnarray}
\left|J^P=\frac{3}{2}^-,2\right\rangle=+\sqrt{\frac{5}{6}}\left|^2P_{\lambda}\frac{3}{2}^-\right\rangle+\sqrt{\frac{1}{6}}\left|^4P_{\lambda}\frac{3}{2}^-\right\rangle,\\
\left|J^P=\frac{3}{2}^-,1\right\rangle=-\sqrt{\frac{1}{6}}\left|^2P_{\lambda}\frac{3}{2}^-\right\rangle+\sqrt{\frac{5}{6}}\left|^4P_{\lambda}\frac{3}{2}^-\right\rangle.
\end{eqnarray}
According to the predicted masses collected in Table~\ref{mass-spectrum}, the masses of the two $J^P=3/2^-$ states are in the region of $M\simeq(6050-6140)$ MeV. Considering the uncertainties of the predicted masses, we plot the variations of partial decay widths as functions of the masses of the $\Sigma_b|J^P=\frac{3}{2}^-,2\rangle$ and $\Sigma_b|J^P=\frac{3}{2}^-,1\rangle$ states in Fig.~\ref{fig-mass} as well. It is found that the decay properties of the two $J^P=3/2^-$ states are sensitive to the masses. For the $\Sigma_b|J^P=\frac{3}{2}^-,2\rangle$ state, the decay width is about $\Gamma\sim(23-68)$ MeV with the mass in the region of $M=(6050-6140)$ MeV, and it mainly decays into $\Lambda_b\pi$.

Considering the $\Sigma_b(6097)$ as the $\Sigma_b|J^P=\frac{3}{2}^-,2\rangle$ state,
we find it has a relatively broad width of (see Table~\ref{decaywidth})
\begin{eqnarray}
\Gamma \simeq42~\text{MeV},
\end{eqnarray}
which is slightly larger than the observed width of $\Gamma_{\text{exp.}}\simeq 30$ MeV at LHCb~\cite{Aaij:2018tnn}.
Our predicted branching fraction of the dominant decay mode $\Sigma_b\pi$ is
\begin{eqnarray}
\mathcal{B}[\Sigma_b(6097)\to \Lambda_b\pi]\simeq 84\%,
\end{eqnarray}
which is consistent with the fact that the $\Sigma_b(6097)$ resonance was first
observed in the final state $\Lambda_b\pi$~\cite{Aaij:2018tnn}.
Meanwhile, the decay rate of $\Sigma_b(6097)$ into $\Sigma_b\pi$ is considerable, and the predicted branching fraction is
\begin{eqnarray}
\mathcal{B}[\Sigma_b(6097)\to \Sigma_b\pi]\simeq 7.8\%.
\end{eqnarray}
The decay mode $\Sigma_b\pi$ may be observed in future experiments.
As a whole, the state $\Sigma_b(6097)$ is a good candidate of the $J^P=3/2^-$ state $\Sigma_b|J^P=\frac{3}{2}^-,2\rangle$.

For the other $J^P=3/2^-$ state $\Sigma_b|J^P=\frac{3}{2}^-,1\rangle$,
the decay width is predicted to be $\Gamma\sim(20-37)$ MeV with the masses
varied in the range what we considered in the present work, and the
$\Sigma^*_b\pi$ decay channel almost saturates its total decay widths.
If the $\Sigma_b(6097)$ is the $\Sigma_b|J^P=\frac{3}{2}^-,2\rangle$ indeed,
the mass of $\Sigma_b|J^P=\frac{3}{2}^-,1\rangle$ is predicted to be 6087 MeV.
Taking the mass of $\Sigma_b|J^P=\frac{3}{2}^-,1\rangle$ with $M=6087$ MeV,
we obtain the decay width
\begin{eqnarray}
\Gamma \simeq 28~\text{MeV}.
\end{eqnarray}
The dominant decay mode is $\Sigma^*_b\pi$ with the branching fraction
\begin{eqnarray}
\mathcal{B}[\Sigma_b|J^P=3/2^-, 1\rangle\rightarrow \Sigma^*_b\pi]\simeq 95\%.
\end{eqnarray}
To establish the $\Sigma_b|J^P=\frac{3}{2}^-,1\rangle$ state, the $\Sigma^*_b\pi$
is worth to observing in future experiments.
Since the $\Sigma_b^*$ dominantly decay into the $\Lambda_b\pi$ channel, thus,
the $\Sigma_b|J^P=\frac{3}{2}^-,1\rangle$ state should also have a large decay rate into the $\Lambda_b\pi\pi$ channel
via the intermediate $\Sigma^*_b\pi$ process.

Considering the mass of $\Sigma_b|J^P=\frac{3}{2}^-,1\rangle$, $M\simeq6087$ MeV,
is very to that of the broad structure $\Sigma_b(6072)$ observed in the $\Lambda_b\pi\pi$
invariant mass spectrum~\cite{Sirunyan:2020gtz,Aaij:2020rkw}, thus, we may
conclude that the $\Sigma_b|J^P=\frac{3}{2}^-,1\rangle$ state should contribute
to the structure $\Sigma_b(6072)$ as well. Then, we can understand why we
cannot explain width of the $\Sigma_b(6072)$ with the $\Sigma_b|J^P=\frac{3}{2}^-,1\rangle$
state only. As a conclusion, the broad structure may be caused by two highly
overlapping states $\Sigma_b|J^P=\frac{1}{2}^-,1\rangle$ and
$\Sigma_b|J^P=\frac{3}{2}^-,1\rangle$ with dominant decay channels $\Sigma_b\pi$
and $\Sigma^*_b\pi$, respectively.

It should be pointed out that
in Refs.~\cite{Aaij:2020rkw,Arifi:2020yfp} the $\Sigma_b(6072)$ was suggested to be the
$\Lambda_b(2S)$ resonance. However, as the $\Lambda_b(2S)$ assignment, the width
of $\Sigma_b(6072)$ is predicted to be very narrow ($\Gamma\sim 1$ MeV) with our chiral quark model~\cite{Wang:2019uaj}.
Within the quark pair creation model, the authors in Ref.~\cite{Liang:2019aag} obtained that the
total decay widths of $\Lambda_b(2S)$ was about $\sim$12 MeV, which was also inconsistent with the
measured widths $\Gamma=72\pm11\pm2$ from the LHCb~\cite{Aaij:2020rkw}.

\subsection{$J^P=5/2^-$ state}

There is only one $\lambda$-mode $1P$ wave $\Sigma_b$ state with $J^P=5/2^-$, which is no difference between the $j$-$j$ coupling scheme and $L$-$S$ coupling scheme, namely
\begin{eqnarray}
\left|J^P=\frac{5}{2}^-,2\right\rangle =\left|^4P_{\lambda}\frac{5}{2}^-\right\rangle.
\end{eqnarray}
The predicted mass of the state $\Sigma_b|J^P=\frac{5}{2}^-,2\rangle$ is listed in Table~\ref{mass-spectrum}. From the table, the theoretical mass are about $M\simeq(6050-6140)$ MeV. The predicted mass of the $\Sigma_b|J^P=\frac{5}{2}^-,2\rangle$ state certainly has a large uncertainty, which may bring uncertainties to the theoretical results. To investigate this effect, we plot the decay properties of this state as a function of the mass in Fig.~\ref{fig-mass} as well. Our results indicate that the total decay width of $\Sigma_b|J^P=\frac{5}{2}^-,2\rangle$ is about $\Gamma\sim(26-68)$ MeV with the mass varied in the region of $M=(6050-6140)$ MeV, and its strong decays are governed by the $\Lambda_b\pi$ channel in the whole mass region considered in the present work.

According to our previous analysis in Sec.~\ref{massspecrum}, the
mass of $\Sigma_b|J^P=\frac{5}{2}^-,2\rangle$ may be 6107 MeV.
Taking this mass for $\Sigma_b|J^P=\frac{5}{2}^-,2\rangle$, we predict its width to be
\begin{eqnarray}
\Gamma  \simeq50~\text{MeV},
\end{eqnarray}
and the predicted branching ratio is
\begin{eqnarray}
\mathcal{B}[\Sigma_b|J^P=5/2^-, 2\rangle\rightarrow \Lambda_b\pi]\simeq 78\%.
\end{eqnarray}
Meanwhile, the decay rate of $\Sigma_b|J^P=\frac{5}{2}^-,2\rangle$ into $\Sigma^*_b\pi$ is considerable with the predicted branching fraction
\begin{eqnarray}
\mathcal{B}[\Sigma_b|J^P=5/2^-, 2\rangle\rightarrow \Sigma^*_b\pi]\simeq 14\%.
\end{eqnarray}

We notice that the measured natures of the new peak $\Sigma_b(6097)$ observed in the $\Lambda_b\pi$~\cite{Aaij:2018tnn} are good consistent with the predicted properties of the state $\Sigma_b|J^P=\frac{5}{2}^-,2\rangle$. Thus, the new peak $\Sigma_b(6097)$ can be explained as the $J^P=5/2^-$ state $\Sigma_b|J^P=\frac{5}{2}^-,2\rangle$ as well. Hence, our predictions indicate that the new peak $\Sigma_b(6097)$ has two candidates, $\Sigma_b|J^P=\frac{3}{2}^-,2\rangle$ and $\Sigma_b|J^P=\frac{5}{2}^-,2\rangle$.
Since the similar decay properties and close masses for the two candidates, it is difficult to clearly distinguish them.
Thus, the new peak $\Sigma_b(6097)$ most likely arises from the overlapping of $\Sigma_b|J^P=\frac{3}{2}^-,2\rangle$ and $\Sigma_b|J^P=\frac{5}{2}^-,2\rangle$.

\section{Summary}

In this paper, we predict a fairly complete $P$-wave singly-heavy baryon
spectrum belonging to $\mathbf{6}_F$ by combining the equal spacing rule with the observations of
the heavy baryon states. Further combining the mass spectrum with the strong decay properties
of the $1P$ wave $\Sigma_b$ states, we give possible interpretations for the
newly observed structures $\Sigma_b(6072)$ and $\Sigma_b(6097)$.

It is found that the newly observed peak $\Sigma_b(6097)$ observed in the $\Lambda_b\pi$ final state is most likely to arise from the overlapping of $\Sigma_b|J^P=\frac{3}{2}^-,2\rangle$ and $\Sigma_b|J^P=\frac{5}{2}^-,2\rangle$. Both $\Sigma_b|J^P=\frac{3}{2}^-,2\rangle$ and
$\Sigma_b|J^P=\frac{5}{2}^-,2\rangle$ dominantly decay into $\Lambda_b\pi$ with a comparable width, $\Gamma\sim 50$ MeV.

The broad structure $\Sigma_b(6072)$ observed in the $\Lambda_b\pi\pi$ final state may arise from the overlapping of  $\Sigma_b|J^P=\frac{1}{2}^-,1\rangle$ and $\Sigma_b|J^P=\frac{3}{2}^-,1\rangle$. The $\Sigma_b|J^P=\frac{1}{2}^-,1\rangle$ and $\Sigma_b|J^P=\frac{3}{2}^-,1\rangle$ state dominantly decay into $\Sigma_b\pi$ and $\Sigma^*_b\pi$, respectively,
with a comparable width, $\Gamma\sim 30$ MeV.

The $\Sigma_b|J^P=\frac{1}{2}^-,0\rangle$, is most likely to be a quite narrow state with a total decay width of $\Gamma\sim10$ MeV. The $\Lambda_b\pi$ decay channel almost saturates its total decay widths. Considering the predicted width of $\Sigma_b|J^P=\frac{1}{2}^-,0\rangle$ being narrow, this state might be observed in the $\Lambda_b\pi$ channel when enough data are accumulated in experiments.

\section*{Acknowledgements }
This work is supported by the National Natural
Science Foundation of China under Grants No. 11775078, No. 11947048 and  No. U1832173.



\begin{thebibliography}{99}

\bibitem{Aaij:2017nav}
  R.~Aaij {\it et al.} [LHCb Collaboration],
  Observation of five new narrow $\Omega_c^0$ states decaying to $\Xi_c^+ K^-$,
  Phys.\ Rev.\ Lett.\  {\bf 118}, 182001 (2017).

\bibitem{Yelton:2017qxg}
  J.~Yelton {\it et al.} [Belle Collaboration],
  Observation of Excited $\Omega_c$ Charmed Baryons in $e^+e^-$ Collisions,
  Phys.\ Rev.\ D {\bf 97}, 051102 (2018).

\bibitem{Aaij:2020yyt}
  R.~Aaij {\it et al.} [LHCb Collaboration],
  Observation of new $\Xi_c^0$ baryons decaying to $\Lambda_c^+ K^-$,
  arXiv:2003.13649 [hep-ex].


\bibitem{Aaij:2018tnn}
  R.~Aaij {\it et al.} [LHCb Collaboration],
  Observation of Two Resonances in the $\Lambda_b^0 \pi^\pm$ Systems and Precise Measurement of $\Sigma_b^\pm$ and $\Sigma_b^{*\pm}$ properties,
  Phys.\ Rev.\ Lett.\  {\bf 122}, 012001 (2019).


\bibitem{Aaij:2018yqz}
  R.~Aaij {\it et al.} [LHCb Collaboration],
  Observation of a new $\Xi_b^-$ resonance,
  Phys.\ Rev.\ Lett.\  {\bf 121}, 072002 (2018).

\bibitem{Aaij:2020cex}
  R.~Aaij {\it et al.} [LHCb Collaboration],
  First observation of excited $\Omega_b^-$ states,
  arXiv:2001.00851 [hep-ex].


\bibitem{Capstick:2000qj}
  S.~Capstick and W.~Roberts,
  Quark models of baryon masses and decays,
  Prog.\ Part.\ Nucl.\ Phys.\   45, S241 (2000).

\bibitem{Ebert:2005xj}
  D.~Ebert, R.~N.~Faustov and V.~O.~Galkin,
  Masses of heavy baryons in the relativistic quark model,
  Phys.\ Rev.\ D 72, 034026 (2005).


\bibitem{Valcarce:2008dr}
  A.~Valcarce, H.~Garcilazo and J.~Vijande,
  Towards an understanding of heavy baryon spectroscopy,
  Eur.\ Phys.\ J.\ A  37, 217 (2008).


\bibitem{Karliner:2008sv}
  M.~Karliner, B.~Keren-Zur, H.~J.~Lipkin and J.~L.~Rosner,
  The Quark Model and $b$ Baryons,
  Annals Phys.\   324, 2 (2009).

\bibitem{Chen:2014nyo}
  B.~Chen, K.~W.~Wei and A.~Zhang,
  Assignments of $\Lambda_Q$ and $\Xi_Q$ baryons in the heavy quark-light diquark picture,
  Eur.\ Phys.\ J.\ A  51, 82 (2015).

\bibitem{Wang:2010it}
  Z.~G.~Wang,
  Analysis of the ${1/2^-}$ and ${3/2^-}$ heavy and doubly heavy baryon states with QCD sum rules,
  Eur.\ Phys.\ J.\ A  47, 81 (2011).

\bibitem{Jia:2019bkr}
  D.~Jia, W.~N.~Liu and A.~Hosaka,
  Regge behaviors in orbitally excited spectroscopy of charmed and bottom baryons,
  Phys.\ Rev.\ D  101, 034016 (2020).


\bibitem{Capstick:1986bm}
  S.~Capstick and N.~Isgur,
  Baryons in a Relativized Quark Model with Chromodynamics,
  Phys.\ Rev.\ D  34, 2809 (1986).

\bibitem{Yamaguchi:2014era}
  Y.~Yamaguchi, S.~Ohkoda, A.~Hosaka, T.~Hyodo and S.~Yasui,
  Heavy quark symmetry in multihadron systems,
  Phys.\ Rev.\ D  91, 034034 (2015).

\bibitem{Yoshida:2015tia}
  T.~Yoshida, E.~Hiyama, A.~Hosaka, M.~Oka and K.~Sadato,
  Spectrum of heavy baryons in the quark model,
  Phys.\ Rev.\ D  92, 114029 (2015).

\bibitem{Thakkar:2016dna}
  K.~Thakkar, Z.~Shah, A.~K.~Rai and P.~C.~Vinodkumar,
  Excited State Mass spectra and Regge trajectories of Bottom Baryons,
  Nucl.\ Phys.\ A  965, 57 (2017).

\bibitem{Ebert:2011kk}
  D.~Ebert, R.~N.~Faustov and V.~O.~Galkin,
  Spectroscopy and Regge trajectories of heavy baryons in the relativistic quark-diquark picture,
  Phys.\ Rev.\ D  84, 014025 (2011).

\bibitem{Yang:2017qan}
  G.~Yang, J.~Ping and J.~Segovia,
  The $S$- and $P$-Wave Low-Lying Baryons in the Chiral Quark Model,
  Few Body Syst.\   59, 113 (2018).

\bibitem{Roberts:2007ni}
  W.~Roberts and M.~Pervin,
  Heavy baryons in a quark model,
  Int.\ J.\ Mod.\ Phys.\ A 23, 2817 (2008).

\bibitem{Garcilazo:2007eh}
  H.~Garcilazo, J.~Vijande and A.~Valcarce,
  Faddeev study of heavy baryon spectroscopy,
  J.\ Phys.\ G 34, 961 (2007).

\bibitem{Ebert:2007nw}
  D.~Ebert, R.~N.~Faustov and V.~O.~Galkin,
  Masses of excited heavy baryons in the relativistic quark model,
  Phys.\ Lett.\ B  659, 612 (2008).

\bibitem{Wei:2016jyk}
  K.~W.~Wei, B.~Chen, N.~Liu, Q.~Q.~Wang and X.~H.~Guo,
  Spectroscopy of singly, doubly, and triply bottom baryons,
  Phys.\ Rev.\ D 95, 116005 (2017).

\bibitem{Mao:2015gya}
  Q.~Mao, H.~X.~Chen, W.~Chen, A.~Hosaka, X.~Liu and S.~L.~Zhu,
  QCD sum rule calculation for $P$-wave bottom baryons,
  Phys.\ Rev.\ D  92, 114007 (2015).

\bibitem{Chen:2018vuc}
  B.~Chen and X.~Liu,
  Assigning the newly reported $\Sigma_b(6097)$ as a $P$-wave excited state and predicting its partners,
  Phys.\ Rev.\ D {\bf 98}, 074032 (2018).


\bibitem{Wang:2020gkn}
  K.~L.~Wang, L.~Y.~Xiao and X.~H.~Zhong,
  Understanding the newly observed $\Xi_c^0$ states through their decays,
  arXiv:2004.03221 [hep-ph].

\bibitem{Xiao:2020oif}
  L.~Y.~Xiao, K.~L.~Wang, M.~S.~Liu and X.~H.~Zhong,
  Possible interpretation of the newly observed $\Omega_b$ states,
  Eur.\ Phys.\ J.\ C  80, 279 (2020).

\bibitem{Wang:2017kfr}
  K.~L.~Wang, Y.~X.~Yao, X.~H.~Zhong and Q.~Zhao,
  Strong and radiative decays of the low-lying $S$- and $P$-wave singly heavy baryons,
  Phys.\ Rev.\ D  96, 116016 (2017).

\bibitem{Wang:2018fjm}
  K.~L.~Wang, Q.~F.~L$\ddot{\text{u}}$ and X.~H.~Zhong,
  Interpretation of the newly observed $\Sigma_b(6097)^{\pm}$ and $\Xi_b(6227)^-$ states as the $P$-wave bottom baryons,
  Phys.\ Rev.\ D  99, 014011 (2019).


\bibitem{Wang:2017hej}
  K.~L.~Wang, L.~Y.~Xiao, X.~H.~Zhong and Q.~Zhao,
  Understanding the newly observed $\Omega_c$ states through their decays,
  Phys.\ Rev.\ D  95, 116010 (2017).

\bibitem{Sirunyan:2020gtz}
  A.~M.~Sirunyan {\it et al.} [CMS Collaboration],
  Study of excited $\Lambda_\mathrm{b}^0$ states decaying to $\Lambda_\mathrm{b}^0\pi^+\pi^-$ in proton-proton collisions at $\sqrt{s}=$ 13 TeV,
  Phys.\ Lett.\ B  803, 135345 (2020).

\bibitem{Aaij:2020rkw}
  R.~Aaij {\it et al.} [LHCb Collaboration],
  Observation of a new baryon state in the $\Lambda_b^0\pi^+\pi^-$ mass spectrum,
  arXiv:2002.05112 [hep-ex].


\bibitem{Arifi:2020yfp}
  A.~J.~Arifi, H.~Nagahiro, A.~Hosaka and K.~Tanida,
  Roper-like resonances with various flavor contents and their two-pion emission decays,
  arXiv:2004.07423 [hep-ph].


\bibitem{GellMann:1962xb}
  M.~Gell-Mann,
  Symmetries of baryons and mesons,
  Phys. Rev. 125, 1067 (1962).

\bibitem{Okubo:1961jc}
  S.~Okubo,
  Note on unitary symmetry in strong interactions,
  Prog. Theor. Phys.  27, 949 (1962).


\bibitem{Zhong:2007gp}
  X.~H.~Zhong and Q.~Zhao,
  Charmed baryon strong decays in a chiral quark model,
  Phys.\ Rev.\ D  77, 074008 (2008).

\bibitem{Liu:2012sj}
  L.~H.~Liu, L.~Y.~Xiao and X.~H.~Zhong,
  Charm-strange baryon strong decays in a chiral quark model,
  Phys.\ Rev.\ D  86, 034024 (2012).


\bibitem{Yao:2018jmc}
  Y.~X.~Yao, K.~L.~Wang and X.~H.~Zhong,
  Strong and radiative decays of the low-lying $D$-wave singly heavy baryons,
  Phys.\ Rev.\ D  98, 076015 (2018).

\bibitem{Xiao:2017udy}
  L.~Y.~Xiao, K.~L.~Wang, Q.~f.~Lu, X.~H.~Zhong and S.~L.~Zhu,
  Strong and radiative decays of the doubly charmed baryons,
  Phys.\ Rev.\ D 96, 094005 (2017).

\bibitem{Xiao:2013xi}
  L.~Y.~Xiao and X.~H.~Zhong,
  $\Xi$ baryon strong decays in a chiral quark model,
  Phys.\ Rev.\ D  87, 094002 (2013).


\bibitem{Zhong:2008kd}
  X.~h.~Zhong and Q.~Zhao,
  Strong decays of heavy-light mesons in a chiral quark model,
  Phys.\ Rev.\ D  78, 014029 (2008).

\bibitem{Zhong:2009sk}
  X.~H.~Zhong and Q.~Zhao,
  Strong decays of newly observed D(sJ) states in a constituent quark model with effective Lagrangians,
  Phys.\ Rev.\ D 81, 014031 (2010).

\bibitem{Zhong:2010vq}
  X.~H.~Zhong,
  Strong decays of the newly observed $D(2550)$, $D(2600)$, $D(2750)$, and $D(2760)$,
  Phys.\ Rev.\ D  82, 114014 (2010).

\bibitem{Xiao:2014ura}
  L.~Y.~Xiao and X.~H.~Zhong,
  Strong decays of higher excited heavy-light mesons in a chiral quark model,
  Phys.\ Rev.\ D 90, 074029 (2014).


\bibitem{Wang:2019uaj}
  K.~L.~Wang, Q.~F.~L$\ddot{\text{u}}$ and X.~H.~Zhong,
  Interpretation of the newly observed $\Lambda_b(6146)^{0}$ and $\Lambda_b(6152)^0$ states in a chiral quark model,
  Phys.\ Rev.\ D 100, 114035 (2019).



\bibitem{Liang:2019aag}
  W.~Liang, Q.~F.~L¨¹ and X.~H.~Zhong,
  Canonical interpretation of the newly observed $\Lambda_b(6146)^0$ and $\Lambda_b(6152)^0$ via strong decay behaviors,
  Phys.\ Rev.\ D 100, 054013 (2019).




\end{thebibliography}
\end{document}